\documentclass[prl,twocolumn,showpacs]{revtex4}

\usepackage{psfig}

\renewcommand{\vec}[1]{{\rm\bf #1}}
\newcommand{\ep}{\epsilon}
\newcommand{\bra}[1]{\left\langle{#1}\right|}
\newcommand{\ket}[1]{\left|{#1}\right\rangle}
\newcommand{\ScProd}[2]{\left\langle{#1}\!\right.\left|{#2}\right\rangle}
\newcommand{\unimat}{\hat{\bf 1}}

\newcommand{\Stlin}{\mathop{\widehat{\mathrm{St}}}}

\begin{document}

\title{Hopping between localized Floquet states in periodically driven
quantum dots}
\author{D.~M.~Basko}
\email{basko@ictp.trieste.it} \affiliation{The Abdus Salam
International Centre for Theoretical Physics, Strada Costiera 11,
34100 Trieste, Italy}

\date{\today}
\begin{abstract}
The dynamic localization in energy space -- suppression of the
absorption of energy from an external microwave field due to
quantum interference -- was analyzed recently for a closed quantum
dot in the absence of electron-electron interactions. Here a weak
interaction is shown to lead to a finite absorption and heating,
which may be viewed as hopping between localized Floquet states.
The heating rate grows together with the electronic temperature,
eventually destroying the localization.
\end{abstract}


\maketitle

{\em Introduction.}--- The dynamic localization (DL)~\cite{Casati79}
-- suppression of the absorption of energy by a quantum system under
a periodic perturbation -- was studied extensively for quantum
chaotic models~\cite{Fishman,Izrailev}. In a recent
publication~\cite{us} DL was shown to be possible in a solid-state
system -- an ac driven chaotic quantum dot~(QD).
Experiments on such systems started in the last few
years~\cite{Marcus}, which makes the DL in a QD more than just a
theoretical peculiarity. Calculations of Ref.~\cite{us} employed
the simplest random-matrix model for the dot, which neglects
electron-electron interaction. The latter, however, is always
present in a~QD, and studying its effect is the subject of the
present paper.
Many-body effects on the~DL have been studied previously for quite
a different system~\cite{Shepelyansky}, so this issue is also of
a fundamental interest.

Let the single-electron mean level spacing~$\delta$ in the dot
(always assumed to be closed) be {\em the smallest energy scale}
in the problem. Then, if an external time-dependent periodic
perturbation with the frequency~$\omega$ is applied, the total
electronic energy~$E$ in the dot (counted from that of the ground
state) grows linearly with time as described by the Fermi Golden
Rule: $E(t)=\Gamma\omega^2t/(2\delta)\equiv{W}_0t$. The
probability of each single-electron transition per unit time,
denoted by $2\Gamma$, measures the strength of the
perturbation~\cite{perturbation}. The criterion of validity of
the Fermi Golden Rule is $\Gamma\gg\delta$, and $\Gamma\ll\omega$
is also assumed ($\hbar=1$).

This picture (hereafter referred to as Ohmic absorption) is valid
provided that each act of photon absorption by an electron is
independent of the previous ones; however, for a discrete energy
spectrum this turns out not to be the case. After many transitions
the absorption rate decreases due to accumulation of the quantum
interference correction~\cite{us}, so that after a time
$t_*\sim\Gamma/\delta^2$ the absorption is completely suppressed.
This effect was named the dynamic localization in energy
space. It is the consequence of level discreteness: at
$\delta\rightarrow{0}$ it takes longer time for the DL to develop,
and for the continuous spectrum there is no~DL. It should be also
noted that the problem of DL in the energy space turns out to be
analogous to the Anderson localization problem in a
one-dimensional disordered system in the real
space~\cite{Fishman}.

Conduction in a disordered sample in the regime of the strong
Anderson localization occurs via thermally activated hopping
between localized states due to inelastic processes (typically,
phonons)~\cite{Mott}, the activated temperature dependence being
one of the main signatures of the localization. In the QD phonons
are assumed to be frozen out by the external cryostat, hence the
dominant role is played by electron-electron collisions. The
latter are governed by the {\em electronic temperature}~$T$ in the
dot, which is different from the cryostat temperature. In the DL
regime the (effective) electronic temperature
$T_*\sim\Gamma\omega/\delta$ and the localization length in the
energy space are in fact the same thing, which invalidates the
familiar concept of the activated hopping conduction for the DL
problem.

Electron-electron collisions in an equilibrium QD were studied by
Sivan, Imry, and Aronov (SIA)~\cite{Sivan} (see also
Ref.~\cite{Altshuler}). The single-electron relaxation rate was
found to be $\gamma(T)\sim{T}^2\delta/E_{Th}^2$, where $E_{Th}$~is
the Thouless energy. In the following, the hierarchy of scales
$\delta\ll\Gamma\ll\omega\ll{T}_*\ll{E}_{Th}$ is assumed,
$E_{Th}$~being thus {\em the largest energy scale} in the problem.
For the strong DL to take place, the quantum correction should have
enough time to develop, which is governed by a dimensionless
parameter $u\equiv
\gamma(T_*)t_*\sim(\Gamma/\delta)^3(\omega/E_{Th})^2\ll{1}$.

{\em Qualitative picture.}--- The theory to be presented below
results in the following qualitative picture. As the collisions
are rare ($\gamma{t}_*\ll{1}$), the electrons spend most of the
time in the states localized in energy space, having definite
phase relationships. When two electrons collide, the phase memory
is lost for them, and their wave packets start spreading along the
energy axis. They localize again after the time $\sim{t}_*$, in
the meantime spreading by $\sim{T}_*$. Thus, although an
electron-electron collision, strictly speaking, by itself does not
change the total energy of the system, the subsequent ac driven
dynamics leads to a change of the total energy of $\sim{T}_*$~per
collision. The sign of this change is, however, arbitrary, because
a periodic perturbation can equally cause transitions up and down
the spectrum. Only the presence of the filled Fermi sea below
(i.~e., an energy gradient of the electronic distribution
function) makes absorption the preferred direction, which means
that if the electronic temperature $T\gg{T}_*$, the energy
absorbed per collision is $\sim{T}_*^2/T$ rather than~$T_*$. The
effective number of electrons that can participate in a collision
is $\sim{T}/\delta$ (due to the degenerate Fermi statistics).
During the time interval $\sim{1}/\gamma$ each of these electrons
participates in one collision, so the total number of collisions
per unit time is $\sim(T/\delta)\gamma(T)$. This gives the energy
absorption rate:
\begin{equation}\label{WT=}
W\equiv\frac{dE}{dt}\sim
\frac{T_*^2}{T}\,\frac{T}{\delta}\,\gamma(T)\sim
\frac{T^2T_*^2}{E_{Th}^2}\sim\gamma(T_*)E,
\end{equation}
which is the main result of the paper.
The total energy ($E\sim{T}^2/\delta$) is growing
exponentially, however, with the rate being small in the
parameter~$u$. This picture is valid up to~$T$ such that
$\gamma(T)\sim{1}/t_*$, when the strong DL is destroyed. One
can see that at this temperature Eq.~(\ref{WT=}) gives
the Ohmic absorption rate
$W_0\sim{\Gamma}\omega^2/\delta\sim\omega{T}_*$. This happens
for an arbitrarily weak interaction, in contrast to the case
of a kicked nonlinear quantum rotor~\cite{Shepelyansky}.

Eq.~(\ref{WT=}) allows for another simple interpretation. Each
single electron is hit by another one with the average time
interval $1/\gamma$. After a collision the electron spends the
time ${t}_*\ll{1}/\gamma$ spreading its wave packet, and then
it has to wait for the next collision. Thus the absorption rate
of the whole system is given by the simple weighted average:
$W\sim{W}_0t_*\gamma(T)$, which is exactly Eq.~(\ref{WT=}).

\begin{figure}
\psfig{figure=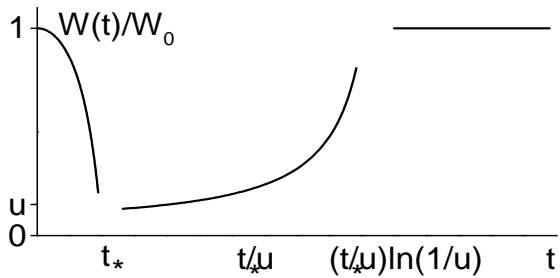,height=4cm,width=8cm}
\caption{\label{absrate:} A sketch of the dependence of the
dimensionless absorption rate $W(t)/W_0$ on time in the case of
strong dynamic localization ($u\equiv\gamma(T_*)t_*\ll{1}$).
Initially the absorption is Ohmic, at $t\sim{t}_*$ it is
suppressed by the DL to a small residual value $W\sim{u}W_0$
(note the algebraic smallness in contrast to an exponential
one for hopping conduction in disordered media).
After a long time $t_*/u=1/\gamma(T_*)$ the exponential heating
becomes noticeable, and after a time $(t_*/u)\ln(1/u)$ the strong
DL is destroyed and $W\sim{W}_0$.}
\end{figure}

{\em Floquet states.}--- It is convenient to start the formal part
of the discussion from some general statements concerning a system
whose hamiltonian $\hat{H}(t)$ is periodic in time with a period
$2\pi/\omega$. According to the Floquet theorem the eigenfunctions
of the time-dependent Schr\"odinger equation have the form
$\ket{\psi_{\alpha}(t)}=e^{-i\lambda_{\alpha}t}\ket{\chi_{\alpha}(t)}$,
where $\ket{\chi_{\alpha}(t)}$ is also periodic in~$t$, and the
range of quasienergies is restricted to
$|\lambda_{\alpha}|<\omega/2$. One can show~\cite{Shirley} that
$\ket{\chi_{\alpha}(t)}$ at any fixed~$t$ form an orthonormal and
complete system:
\begin{equation}\label{orthogonality=}
\ScProd{\chi_{\alpha}(t)}{\chi_{\alpha'}(t)}=
\delta_{\alpha\alpha'}\:,\quad
\sum_{\alpha}\ket{\chi_{\alpha}(t)}\bra{\chi_{\alpha}(t)}=\unimat\:,
\end{equation}
where~$\unimat$ is the unit operator.

If some perturbation $\hat{H}'(t)$ is added to the original
hamiltonian $\hat{H}(t)$, the solution of the Schr\"odinger
equation can be sought in the form
$\ket{\psi(t)}=\sum_{\alpha}C_{\alpha}(t)\ket{\chi_{\alpha}(t)}$,
with the coefficients satisfying
\begin{equation}\label{SchrFloq=}
i\dot{C}_{\alpha}(t)=\lambda_{\alpha}C_{\alpha}(t)+
\sum_{\alpha'}\bra{\chi_{\alpha}(t)}\hat{H}'(t)
\ket{\chi_{\alpha'}(t)}C_{\alpha'}(t).
\end{equation}
Now suppose that some level~$\lambda_{\alpha}$ is resonant with a
continuum of other states~$\lambda_{\alpha'}$, and $\hat{H}'$~does
not depend on time. One can proceed analogously to the textbook
derivation of the Fermi Golden Rule picking up the singular term
in first-order perturbative expression for $|C_{\alpha'}(t)|^2$,
and obtain the transition probability per unit time:
\begin{eqnarray}\label{FGRFloquet=}
dw_{\alpha\rightarrow\alpha'}=2\pi|M_{\alpha'\alpha}|^2
\delta(\lambda_{\alpha'}-\lambda_{\alpha})\,d\alpha',\\ \label{Mtr=}
M_{\alpha'\alpha}=\int\limits_{-\pi/\omega}^{\pi/\omega}
\bra{\chi_{\alpha'}(t)}\hat{H}'\ket{\chi_{\alpha}(t)}
\frac{\omega\,dt}{2\pi}\:.
\end{eqnarray}
As $\lambda_{\alpha'}$~may originate from the part of the spectrum
of~$\hat{H}_0$ which was higher or lower than~$\lambda_{\alpha}$
by several~$\omega$, Eq.~(\ref{FGRFloquet=}) describes
photon-assisted transitions.

{\em Model for the dot.}--- The unperturbed single-particle
hamiltonian~$\hat{H}_0=-\nabla^2/(2m)+v(\vec{r})$ is the usual
one, giving rise to the random-matrix-theory (RMT) spectrum with
the mean level spacing~$\delta$ at energies smaller
than~$E_{Th}$~\cite{Efetov82}. A deterministic form of the
perturbation $\hat{V}(t)$ can lead to the time-dependent
RMT~\cite{Kanzieper}, which is equivalent to another stochastic
model:
\begin{equation}
\hat{V}(t)=V(\vec{r})\,\phi(t)\:,\quad
\overline{V(\vec{r})\,V(\vec{r}')}=
\frac{\Gamma\delta}{\pi}\,L^d\delta(\vec{r}-\vec{r}')
\end{equation}
with $V(\vec{r})$ random and independent of~$v(\vec{r})$,
$\overline{V(\vec{r})}=0$, and $\phi(t)$ being a given periodic
function of time. Periodic functions are expanded in the Fourier
series:
\begin{equation}
\phi(t)=\sum_{s=-\infty}^{\infty}\phi_se^{-is\omega{t}},\quad
\chi_{\alpha}(\vec{r},t)=\sum_{s=-\infty}^{\infty}
\chi_{\alpha}^{(s)}(\vec{r})e^{-is\omega{t}}.
\end{equation}
The harmonics $\phi_s$ are assumed to decrease faster than
$1/s^{3/2}$ to give a finite absorption rate~\cite{us}.
Floquet wave functions satisfy the equation~\cite{Shirley}:
\begin{equation}\label{Schr1D=}
(\hat{H}_0-s\omega)\chi^{(s)}_{\alpha}(\vec{r})+
\sum_{s'}V(\vec{r})\phi_{s-s'}\chi^{(s')}_{\alpha}(\vec{r})=
\lambda_{\alpha}\chi^{(s)}_{\alpha}(\vec{r}).
\end{equation}
No averaging over the disorder has been performed yet.

Eq.~(\ref{Schr1D=}) has a simple and illustrative interpretation.
It describes stationary wave functions for a one-dimensional chain
of coupled granules of the same shape, but with the energy of each
granule~$s$ shifted by~$s\omega$. The coupling broadens each level
of an unperturbed granule by~$\Gamma$.
Due to the condition $\Gamma\gg\delta$ the motion along the chain
is diffusive with the dimensionless diffusion coefficient
$(\Gamma/2)\sum_ss^2|\phi_s|^2$. Due to the condition
$\omega\gg\Gamma$ the spectral correlations within the energy strip
of the width~$\Gamma$ between different granules can be neglected,
and the granules can be considered independent. Wave functions in
such a system have been shown to be localized with the localization
length $\xi\sim\Gamma/\delta\gg{1}$~\cite{Efetov}. Neglecting
fluctuations of the spectrum, one can label the localized states
by the position~$\bar{s}$ (the number of the segment of the
size~$\xi$) and the quasienergy~$\lambda$, introducing the mean
level spacing $\delta_{\xi}\equiv\delta/\xi\sim{1}/t_*$ within one
segment:
\begin{equation}
\sum_{\alpha}\rightarrow
\int\limits_{-\infty}^{\infty}\frac{d\bar{s}}{\xi}
\int\limits_{-\omega/2}^{\omega/2}\frac{d\lambda}{\delta_{\xi}}\:.
\end{equation}

{\em Kinetic equation.}--- As the Floquet states provide a
suitable single-particle basis, one can introduce their occupation
probabilities~$f_{\alpha}$ which are related to the actual energy
distribution function~$f_{\ep}$ via
\begin{equation}
f_{\ep}=\delta\sum_{\alpha,s}f_{\alpha}
\ScProd{\chi_{\alpha}^{(s)}}{\chi_{\alpha}^{(s)}}
\delta(\ep-\lambda_{\alpha}-s\omega)\:.
\end{equation}
In the absence of electron-electron
collisions $f_{\alpha}$~does not depend on time, otherwise one can
write down the collision integral based on the Fermi Golden
Rule~(\ref{FGRFloquet=}):
\begin{widetext}
\begin{equation}
\frac{\partial{f}_{\alpha_1}}{\partial{t}}=
2\pi\sum_{\alpha_{2,3,4}}
\left|M_{\alpha_3\alpha_4\rightarrow\alpha_1\alpha_2}\right|^2
\delta(\lambda_{\alpha_1}+\lambda_{\alpha_2}
-\lambda_{\alpha_3}-\lambda_{\alpha_4})
\left[(1-f_{\alpha_1})(1-f_{\alpha_2})f_{\alpha_3}f_{\alpha_4}
-f_{\alpha_1}f_{\alpha_2}(1-f_{\alpha_3})(1-f_{\alpha_4})\right].
\end{equation}
\end{widetext}
Note that this kinetic equation has a formal stationary
``Fermi-Dirac'' solution
$f_{\alpha}=[e^{\beta(\lambda_{\alpha}-\mu)}+1]^{-1}$. However,
this $\beta$~has nothing to do with the real electronic
temperature, which for this solution is infinite since the states
with all~$\bar{s}$ participate equally. Instead,
$f_{\alpha}$~will be assumed to be independent of~$\lambda$, which
corresponds to the actual distribution function~$f_{\ep}$ to be
smooth on the scale of single-particle energies $\ep\sim{T}_*$,
while the dependence of~$f_{\alpha}$ on~$\bar{s}$
determines~$f_{\ep}$ through $\ep=\bar{s}\omega$.

The matrix element
$M_{\alpha_3\alpha_4\rightarrow\alpha_1\alpha_2}$ originating from
a two-particle interaction $U(\vec{r}_1,\vec{r}_2)$ can be estimated
from Eq.~(\ref{Mtr=}) by taking non-interacting two-particle Floquet
wave functions to be direct products of one-particle ones. It has
the obvious dependence on the positions:
\begin{equation}
M_{\alpha_3\alpha_4\rightarrow\alpha_1\alpha_2}\sim
M_{\lambda_3\lambda_4\rightarrow\lambda_1\lambda_2}
e^{-|\bar{s}_1+\bar{s}_2-\bar{s}_3-\bar{s}_4|/\xi},
\end{equation}
which comes from the overlap of the localized wave functions, the
preexponential factor representing the matrix element for
$|\bar{s}_1+\bar{s}_2-\bar{s}_3-\bar{s}_4|\sim\xi$. To estimate it
one can use the fact that the particle motion within one
localization segment is diffusive. The starting point is
\begin{widetext}
\begin{eqnarray*}
\left|M_{\lambda_3\lambda_4\rightarrow\lambda_1\lambda_2}\right|^2&=&
\sum_{s_1,s_2,s_2'}\sum_{s_3,s_4,s_4'}
\int{d}^d\vec{r}_1\,\ldots\,{d}^d\vec{r}_4\, 
[\chi^{(s_1)}_{\lambda_1}(\vec{r}_1)\,
\chi^{(s_2)}_{\lambda_2}(\vec{r}_2)]^*\,
U(\vec{r}_1,\vec{r}_2)\,\chi^{(s_2')}_{\lambda_3}(\vec{r}_2)\,
\chi_{\lambda_4}^{(s_1+s_2-s_2')}(\vec{r}_1) \times \\ && \times
\chi_{\lambda_1}^{(s_3)}(\vec{r}_3)\,
\chi_{\lambda_2}^{(s_4)}(\vec{r}_4)\, U(\vec{r}_3,\vec{r}_4)\,
[\chi_{\lambda_3}^{(s_4')}(\vec{r}_4)\,
\chi_{\lambda_4}^{(s_3+s_4-s_4')}(\vec{r}_3)=\\
&=&\left(\frac{\delta_{\xi}}{2\pi}\right)^4 \sum_{\{s_i\}}
\int\{d^d\vec{r}_i\}\,U(\vec{r}_1,\vec{r}_2)\, 
(G^R-G^A)(\vec{r}_2,s_2';\vec{r}_4,s_4';\lambda_3)\,
(G^R-G^A)(\vec{r}_4,s_4;\vec{r}_2,s_2;\lambda_2) \times \\ &&
\times U(\vec{r}_3,\vec{r}_4)\,
(G^R-G^A)(\vec{r}_3,s_3;\vec{r}_1,s_1;\lambda_1)\,
(G^R-G^A)(\vec{r}_1,s_1+s_2-s_2';\vec{r}_3,s_3+s_4-s_4';\lambda_4),
\end{eqnarray*}
\end{widetext}
the retarded and advanced Green's functions of the Schr\"odinger
equation~(\ref{Schr1D=}) introduced in the usual way:
\begin{equation}
G^{R,A}(\vec{r},s;\vec{r}',s';\lambda)=
\sum_{\lambda'}\frac{\chi_{\lambda'}^{(s)}(\vec{r})\,
[\chi_{\lambda'}^{(s')}(\vec{r}')]^*}{\lambda-\lambda'\pm{io}}\:,
\end{equation}
where the quasienergy~$\lambda'$ labels the states within the
segment. It is technically simpler to consider the electron motion
in the dot to be diffusive rather than ballistic (the final
results are the same~\cite{Glazman}). When averaging the products
of Green's functions, one introduces the diffusion ladder dressed
by the coupling~$V$ to the second order. The derivation, though
cumbersome, is quite standard~\cite{diffusion}. The part of the
screened Coulomb interaction, responsible for collisions, has the
form~\cite{Glazman}:
\begin{equation}
U(\vec{r}_1,\vec{r}_2)=\frac{L^d\delta}{2}\,\delta(\vec{r}_1-\vec{r}_2)\:,
\end{equation}
and the final result is:
\begin{equation}
\overline{\left|M_{\lambda_3\lambda_4\rightarrow\lambda_1\lambda_2}\right|^2}
=\frac{\delta^4}{(2\pi)^2\xi}\sum_{j>0}\frac{1}{\Lambda_j^2}
\sim\frac{\delta^4}{\xi{E}_{Th}^2}\:.
\end{equation}
Here $\Lambda_j$ are the eigenvalues of the diffusion operator in
the dot in the usual sense.

Passing to the distribution function~$f_{\ep}$ as mentioned above
(integration over $\lambda_2,\lambda_3,\lambda_4$ in the collision
integral gives a factor $\omega^2/\delta_{\xi}^3$), one arrives at
the kinetic equation:
\begin{widetext}
\begin{equation}\label{kinetic=}
\frac{\partial{f}_{\ep_1}}{\partial{t}}=\mathrm{St}[f]\equiv
\int\frac{d\ep_2}{\delta}\,\frac{d\ep_3}{\delta}\,\frac{d\ep_4}{\delta}\,
\mathcal{W}(\ep_1+\ep_2-\ep_3-\ep_4)
\left[(1-f_{\ep_1})(1-f_{\ep_2})f_{\ep_3}f_{\ep_4}
-f_{\ep_1}f_{\ep_2}(1-f_{\ep_3})(1-f_{\ep_4})\right],
\end{equation}
\end{widetext}
\begin{equation}\label{WOmega=}
\mathcal{W}(\Omega)\sim
\frac{\delta^4}{E_{Th}^2}\,\frac{e^{-|\Omega|/T_*}}{T_*},
\quad\mathcal{W}(\Omega)=\mathcal{W}(-\Omega),
\end{equation}
the last property following from the isotropy of the spectrum. If
$\mathcal{W}(\Omega)$ were a $\delta$-function, the collision
integral would be zero for a Fermi-Dirac distribution
$f^T_{\ep}\equiv[e^{\ep/T}+1]^{-1}$. If the characteristic
temperature $T\gg{T}_*$, it is almost a $\delta$-function, so
$f^T_{\ep}$~is a good approximation. However, its temperature can
change with time, so the solution should be sought in the form:
\begin{equation}
f_{\ep}=f^{T}_{\ep}+\tilde{f}_{\ep}\:,\quad
\frac{\partial{f}_{\ep}}{\partial{t}}=
\frac{dT}{dt}\,\frac{\partial{f}^T_{\ep}}{\partial{T}}
+\frac{\partial\tilde{f}_{\ep}}{\partial{t}}\:,
\end{equation}
where $\tilde{f}_{\ep}$~is small in $T_*/T$. $\mathrm{St}[f]$ can
be linearized with respect to~$\tilde{f}$:
$\mathrm{St}[f^T_{\ep}+\tilde{f}_{\ep}]\approx
\mathrm{St}[f^T_{\ep}]+\Stlin\tilde{f}_{\ep}$.
In $\Stlin$ one can approximate
$\mathcal{W}(\Omega)\propto\delta(\Omega)$, then
$\Stlin(\partial{f}^T_{\ep}/\partial{T})=0$, so
$\tilde{f}_{\ep}$~should be taken orthogonal to
$\partial{f}^T_{\ep}/\partial{T}$ (as a function of~$\ep$).
The orthogonal component of the kinetic equation determines then
the single-particle relaxation rate of the Floquet states due to
collisions:
\begin{equation}\label{SIArate=}
\gamma_{\ep}(T)=\frac{\pi^2T^2+\ep^2}{2\delta^2}
\int\mathcal{W}(\Omega)\,\frac{d\Omega}{\delta}\sim
\delta\,\frac{\pi^2T^2+\ep^2}{E_{Th}^2}\:.
\end{equation}
This coincides with the SIA expression~\cite{Sivan}, $T_*$~having
dropped out~\cite{diffusion}. The component of the kinetic equation
along $\partial{f}^T_{\ep}/\partial{T}$ determines the heating:
\begin{equation}
\frac{dT}{dt}\,\frac{\partial{f}^T_{\ep}}{\partial{T}}\approx
\frac{15-\pi^2}{4(\pi^2-6)}\,
T\,\frac{\partial{f}^T_{\ep}}{\partial{T}}
\int\mathcal{W}(\Omega)\,\frac{\Omega^2d\Omega}{\delta^3}\:.
\end{equation}
Estimating the integral, one arrives at Eq.~(\ref{WT=}). As a
result, the temperature is growing exponentially as given by
Eq.~(\ref{WT=}) as long as $\gamma_{\ep}(T)\ll{1}/t_*$, otherwise
Fermi Golden Rule~(\ref{FGRFloquet=}) is no longer applicable,
the Floquet states are destroyed, and the absorption becomes Ohmic.

In conclusion, electron-electron collisions were studied in the
regime of the strong dynamic localization in an ac driven
quantum dot. It was shown that in this case the total
electronic energy is effectively not conserved in a collision,
which leads to heating of the electrons in the dot described by
the kinetic equation~(\ref{kinetic=}). As the frequency of
collisions increases with temperature, this heating is exponential
in time, as given by Eq.~(\ref{WT=}). This picture
persists until the strong dynamic localization is destroyed and
the system returns in the Ohmic absorption regime.

The author is greatly indebted to V.~E.~Kravtsov, D.~L.~Maslov,
B.~N.~Narozhny, and O.~M.~Yevtushenko for helpful discussions
and a critical reading of the paper.


\begin{thebibliography}{99}
\bibitem{perturbation}
The transition rate is proportional to the power of the microwave
and can be estimated as $\Gamma\sim(e\mathcal{E}L)^2/\delta$, where
$\mathcal{E}$~is the amplitude of the electric field in
the dot, $L$~is the dot size, and $1/\delta$ is the density of
states. If one takes into account screening, the field
in the dot is related to the external field~$\mathcal{E}_{ext}$ by
$\mathcal{E}\sim\mathcal{E}_{ext}(\omega/\omega_p)^2$, where
$\omega_p$~is the plasma frequency; at low frequencies the field
penetrates only the screening layer~$1/\kappa$, so one should
substitute $\mathcal{E}L\rightarrow\mathcal{E}_{ext}/\kappa$.
Finally, instead of using microwave one can change the dot shape
modulating the gate voltage.
\end{thebibliography}
\end{document}